\documentclass[12pt]{iopart}

\usepackage[dvips]{graphicx}

\begin{document}

\title{Interplay between superconductivity and magnetism in K-doped EuFe$_{2}$As$_{2}$}

\author{Anupam$^1$, P L Paulose$^2$, H S Jeevan$^3$, C Geibel$^3$, and Z Hossain$^1$}

\address{$^1$ Department of Physics, Indian Institute of Technology, Kanpur 208016, India}
\address{$^2$ Tata Institute of Fundamental Research,Homi Bhabha Road, Mumbai 400005, India}
\address{$^3$ Max Planck Institute for Chemical Physics of Solids, 01187 Dresden, Germany}

\ead{zakir@iitk.ac.in}

\begin{abstract}

   Superconductivity is found in 50 \% K-doped EuFe$_{2}$As$_{2}$ sample below 33 K. Our results of electrical resistivity, magnetic susceptibility and $^{57}$Fe and $^{151}$Eu
M\"{o}ssbauer spectroscopy provide clear evidence that the ordering of the Fe moments observed at 190 K in undoped EuFe$_2$As$_2$ is completely suppressed in our 50\% K doped sample, thus there is no coexistence between the Fe magnetic order and the superconducting state. 
However, short range ordering of the Eu moments is coexisting with the superconducting state below 15 K. A bump in the susceptibility well below T$_c$ as well as a slight broadening of the Fe M\"{o}ssbauer line below 15 K evidence an interplay between the Eu magnetism and the superconducting state.

\end{abstract}

\section{Introduction}
   The recent discovery of superconductivity in quaternary rare earth-transition metal oxypnictides has generated immense activity in the field of condensed matter physics. High superconducting transition temperature of 26 K in F-substituted LaOFeAs \cite{1} was an instant hit as it challenged the monopoly of the copper based oxides as the sole provider of high T$_{c}$ superconductors. Subsequently superconductivity of these oxypnictides could be raised further by replacing La by other rare earths such as Ce, Pr, Nd, Gd, Sm \cite{2,3,4,5,6}. Highest T$_{c}$ achieved so far is 55 K in SmOFeAs \cite{6}. In all these cases the parent compound has a spin density wave transition at high temperature along with the structural transition which is suppressed by suitable doping and this suppression/weakening of SDW transition lead to superconductivity. It was soon realized that a related series of compounds viz. AFe$_{2}$As$_{2}$ (A = Ca, Ba, Sr, Eu) also show similar SDW type anomaly 
 in the resistivity and these could also exhibit superconductivity upon suitable doping or application of external pressure \cite{7,8,9,10}. Indeed superconductivity at high temperature has been observed in Na-doped CaFe$_{2}$As$_{2}$ \cite{11} and K-doped BaFe$_{2}$As$_{2}$, SrFe$_{2}$As$_{2}$ as well as EuFe$_{2}$As$_{2}$ \cite{12,13,14}. Suppression of SDW transition by the application of pressure has also been observed in AFe$_{2}$As$_{2}$ (Ca, Sr, Eu) \cite{15,16,17,18}. Among these EuFe$_{2}$As$_{2}$ has a special place due to the fact that Eu in this compound is in a divalent state and has a large magnetic moment (7 $\mu_{B}$) and hence an ideal candidate to investigate the interplay between superconductivity and magnetism. Many interesting phenomena were discovered from the interplay between local moment magnetism and superconductivity in RRh$_{4}$B$_{4}$ and RNi$_{2}$B$_{2}$C \cite{19,20}. A unique situation was found in HoNi$_{2}$B$_{2}$C where a double reentrance is observed in a narrow range of temperature 
interval which was attributed to the development of a c-axis modulation leading to a ferromagnetic component thus causing superconductivity to weaken/disappear \cite{21}. In this paper our primary concern is to ascertain the magnetic state of Eu-moments and how they affect the superconductivity in this compound.

\section{Experimental details}

We prepared the polycrystalline Eu$_{0.5}$K$_{0.5}$Fe$_{2}$As$_{2}$ using solid state reaction. Starting elements were of high purity. The sample preparation process is similar to that described in reference \cite{14}. Subsequently the sample was annealed at $800\,^{\circ}{\rm C}$ for one week. Phase purity was checked using powder x-ray diffraction and Scanning Electron Microscopy (SEM). Energy dispersive x-ray analysis (EDAX) was used to check the composition of the sample.  Magnetization was measured using SQUID magnetometer (Quantum Design, USA) and electrical resistivity was measured using Physical Properties Measurement System (PPMS, Quantum Design, USA). $^{57}$Fe and $^{151}$Eu M\"{o}ssbauer spectroscopy measurements were performed at various temperatures between 300 K and 4 K using a conventional constant acceleration spectrometer. $^{57}$Co and $^{151}$SmF$_3$ sources were used for the $^{57}$Fe and $^{151}$Eu M\"{o}ssbauer spectroscopy, respectively.


\begin{figure}
\begin{center}

\includegraphics[width=8cm, keepaspectratio]{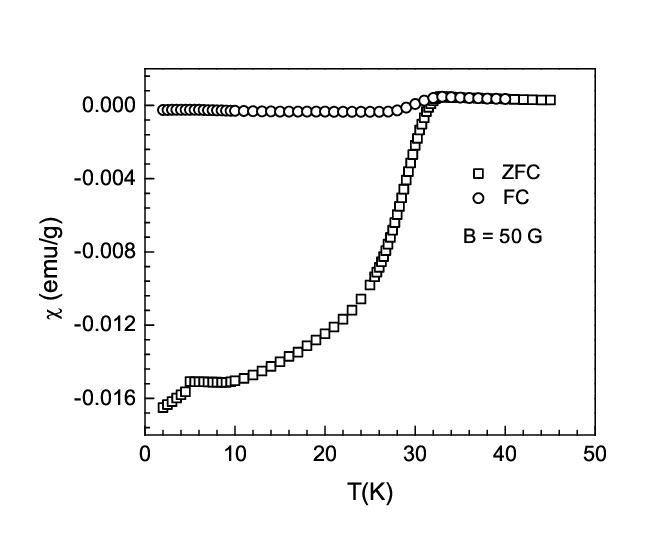}
\caption{\label{Fig 1}Zero Field Cooled (ZFC) and Field Cooled (FC) magnetization as a function of temperature for Eu$_{0.5}$K$_{0.5}$Fe$_{2}$As$_{2}$ measured under applied magnetic field of 50 G.}

\end{center}
\end{figure}
\begin{figure}
\begin{center}

\includegraphics[width=8cm, keepaspectratio]{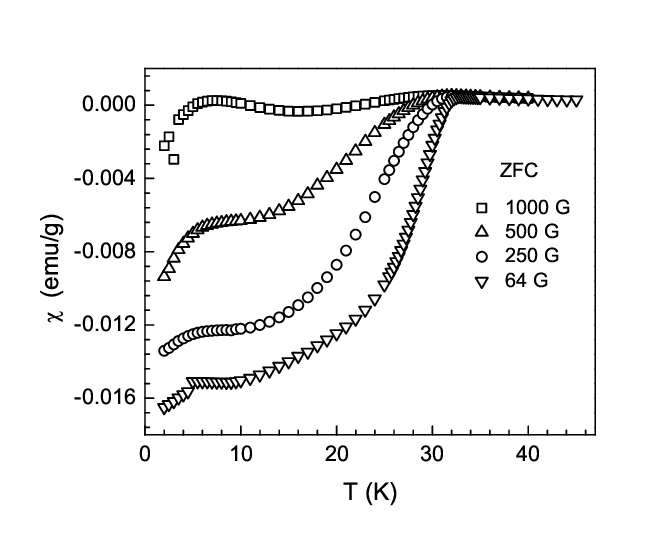}
\caption{\label{Fig 2}Magnetization of Eu$_{0.5}$K$_{0.5}$Fe$_{2}$As$_{2}$ measured under various magnetic fields.}

\end{center}
\end{figure}

\section{Results and discussion}

Polycrystalline Eu$_{0.5}$K$_{0.5}$Fe$_{2}$As$_{2}$ crystallizes in ThCr$_2$Si$_2$ type tetragonal structure (space group I4/mmm) with lattice parameters a = 3.8656 {\AA} and c = 12.962 {\AA} as confirmed by powder x-ray diffraction data. The lattice parameters are in close agreement with the values as reported in \cite{14}. Also the comparison with lattice parameters of  EuFe$_{2}$As$_{2}$ (a = 3.9104 {\AA} and c = 12.1362 {\AA} \cite{10}) shows that there is contraction of the unit cell along a-axis and expansion along c-axis. SEM and EDAX results shows that our sample forms as essentially single phase and homogeneous with avearge composition close to the starting composition of Eu$_{0.5}$K$_{0.5}$Fe$_{2}$As$_{2}$.

          Fig.~1 shows the magnetic susceptibility measured under zero field cooled and field cooled conditions of K-doped EuFe$_{2}$As$_{2}$ for an applied field of 50 G. A clear diamagnetic signal 
corresponding to superconducting transition is observed below 33 K. The superconducting signal under ZFC condition corresponds to that expected for a perfect diamagnetism. Reduced diamagnetic signal under field cooled condition is a hallmark of type II superconductors. In Fig.~2, we plot the magnetic susceptibility under ZFC condition for several values of magnetic fields. In addition to shift of the curves to lower temperature we notice that for 500 G field the magnetic response is similar to that of a reentrant superconductor like HoNi$_{2}$B$_{2}$C. Strength of diamagnetic signal increases initially due to superconductivity, below 15 K the signal shows a tendency to flatten out and then below 7 K diamagnetic signal strength increases again. This kind of behavior was also observed in HoNi$_{2}$B$_{2}$C which was due to destruction/weakening of superconductivity by appearance of ferromagnetic component.

\begin{figure}
\begin{center}

\includegraphics[width=8cm, keepaspectratio]{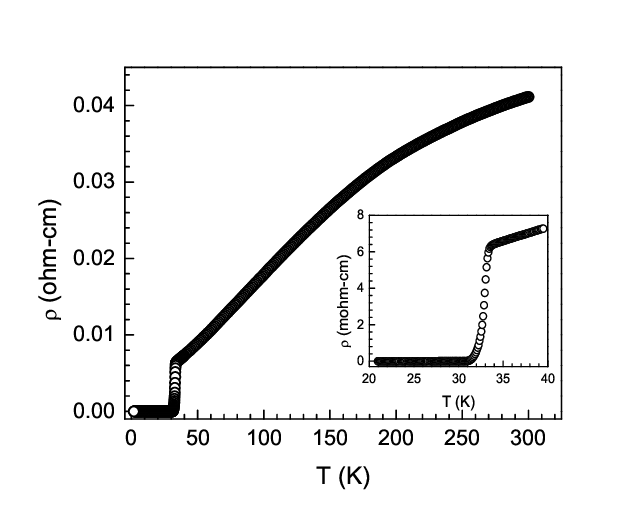}
\caption{\label{Fig 3}Temperature dependence of Electrical resistivity for Eu$_{0.5}$K$_{0.5}$Fe$_{2}$As$_{2}$ at zero field. Inset shows the enlarged view in low temperature range from 20 K to 40 K.}
\end{center}
\end{figure}

\begin{figure}
\begin{center}

\includegraphics[width=8cm, keepaspectratio]{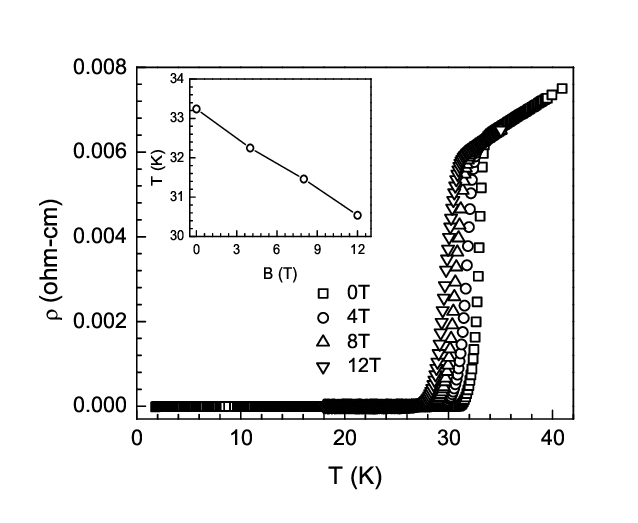}
\caption{\label{Fig 4}Temperature dependence of Electrical resistivity for Eu$_{0.5}$K$_{0.5}$Fe$_{2}$As$_{2}$ at various applied field. Inset shows the B-T phase diagram.}

\end{center}
\end{figure}


\begin{figure}
\begin{center} 

\includegraphics[width=8cm, keepaspectratio]{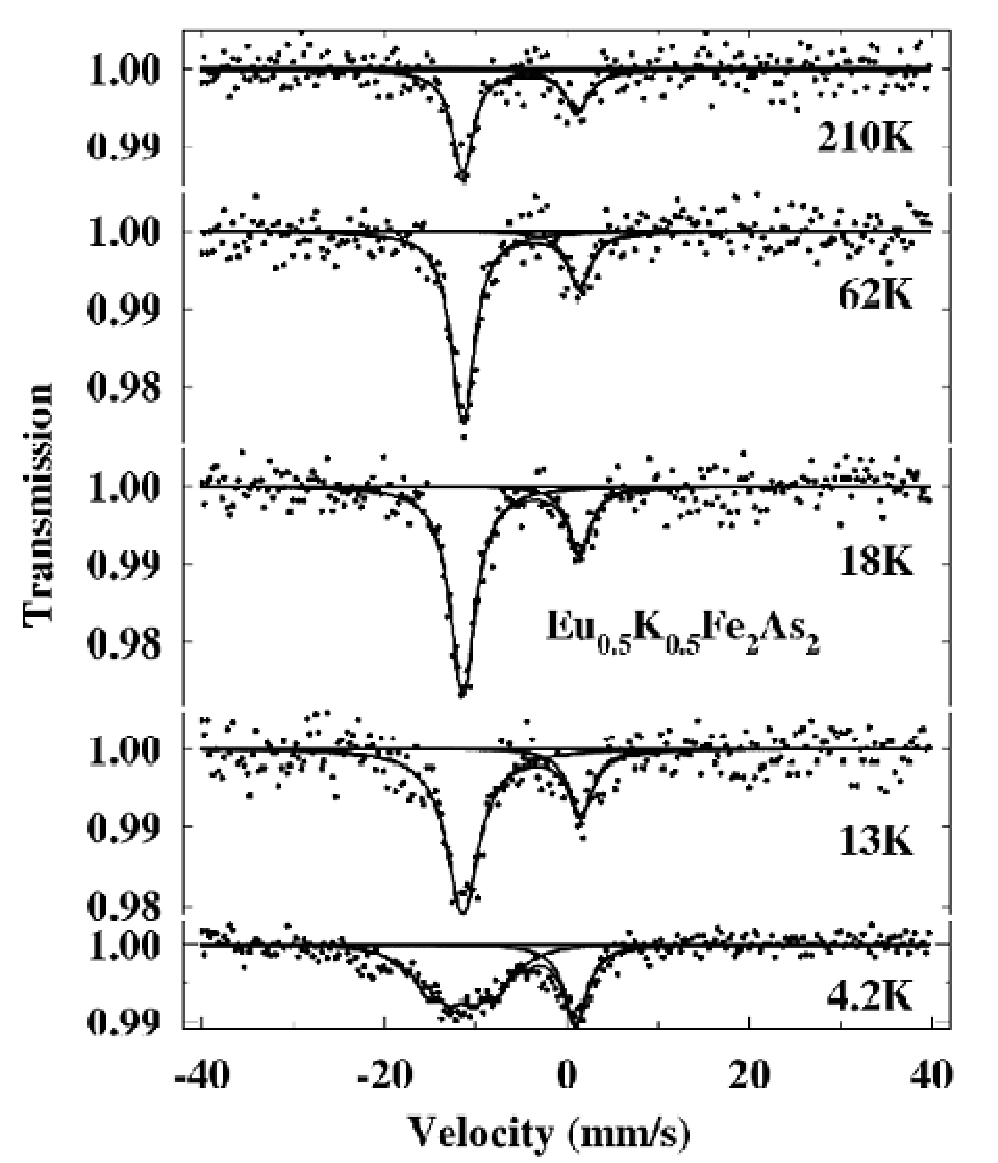}
\caption{\label{Fig 5}$^{151}$Eu M\"{o}ssbauer spectroscopy in the paramagnetic and magnetically ordered state. Note a clear hyperfine splitting of the line at -11.4 mm/s due to magnetic order of Eu$^{2+}$ moments.}
\end{center}
\end{figure}

\begin{figure}
\begin{center}

\includegraphics[width=8cm, keepaspectratio]{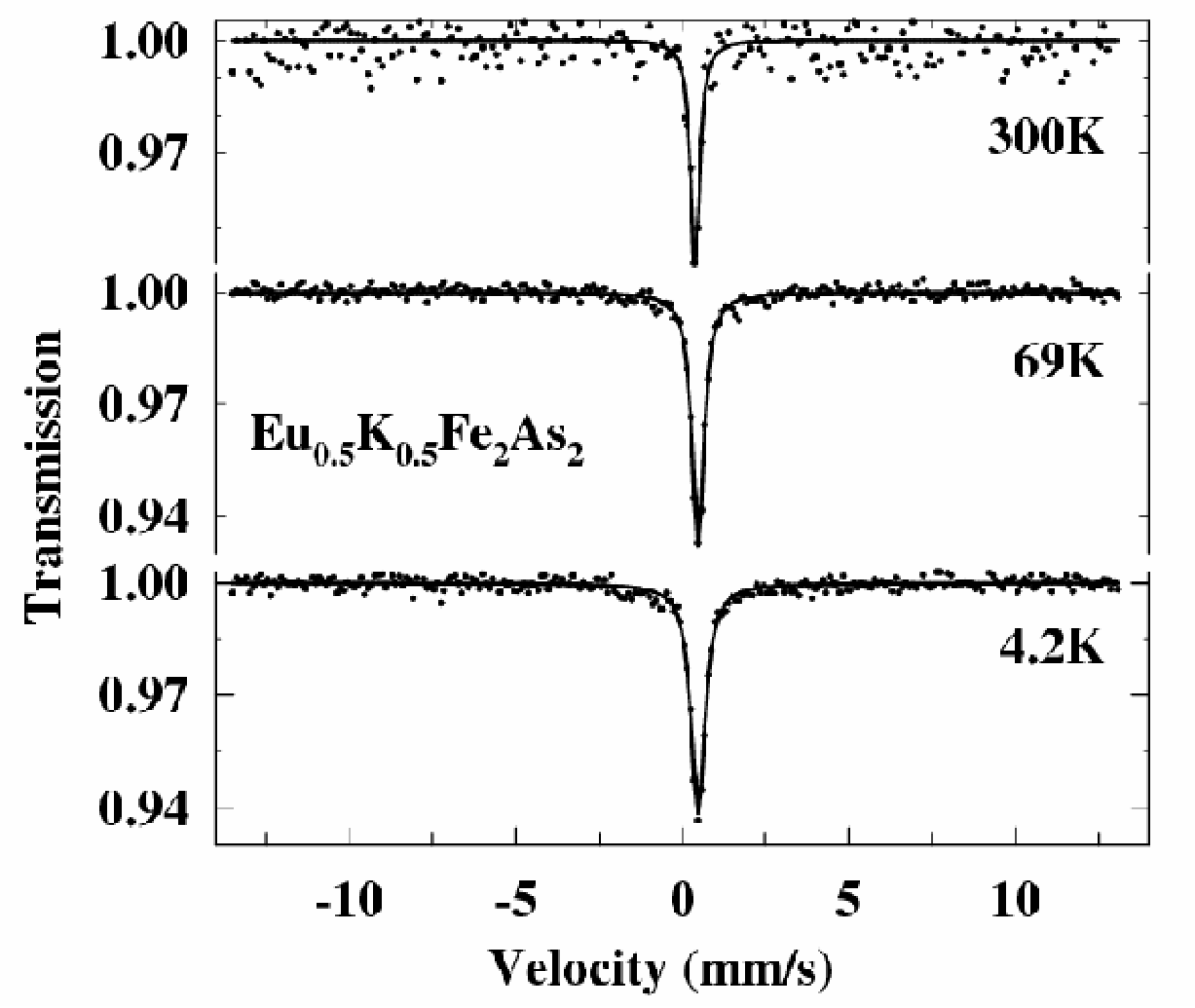}
\caption{\label{Fig 6} $^{57}$Fe M\"{o}ssbauer spectroscopy of Eu$_{0.5}$K$_{0.5}$Fe$_{2}$As$_{2}$ in the paramagnetic state and magnetically ordered state. Only a marginal increase of line width due to magnetic order of Eu-moments is seen below 15 K.}

\end{center}
\end{figure}

The temperature dependence of electrical resistivity for Eu$_{0.5}$K$_{0.5}$Fe$_{2}$As$_{2}$  is shown in Fig.3. Electrical resistivity data exhibits a linear behavior in higher temperature region with an onset of superconducting transition at 34 K while the anomaly due to SDW and structural phase transition as observed in EuFe$_{2}$As$_{2}$ \cite{9, 10} at 190 K is completely suppressed in our 50 \% K-doped EuFe$_{2}$As$_{2}$. The inset in Fig. 3 shows the expanded view in low temperature range 20 K - 40 K. Residual resistivity ratio, RRR ($\rho$ (300 K)/$\rho$ (34 K)) is 6.4, which indicates the good quality of our sample.  Fig. 4 shows the electrical resistivity data $\rho$ (T) in presence of different magnetic fields. As clear from the figure, with increase in magnetic field the onset of superconducting temperature shifts towards lower temperature. Inset in Fig. 4 shows the B-T phase diagram for Eu$_{0.5}$K$_{0.5}$Fe$_{2}$As$_{2}$ based on resistivity measurements. Superconducting transition temperature decreases linearly but slowly with increase in magnetic field. The slope of upper critical field amounts to -0.23 K/T, a value comparable to that observed in Sr$_{0.6}$K$_{0.4}$Fe$_2$As$_2$ \cite{22}.

$^{151}$Eu M\"{o}ssbauer spectroscopy results are shown in Fig. 5. The spectrum at room temperature shows the presence of two single lines with an isomer shifts 1.1 mm/s and -11.4 mm/s. The main line is at -11.4 mm/s which confirm the divalent nature of Eu in this compound. Small intensity line at 1.1 mm/s represents trivalent Eu and this might be due to formation of small amount of trivalent impurity phase during the powdering process for preparing the absorber or during the sample preparation. As the temperature is lowered below 15 K, the line at -11.4 mm/s broadens without showing a well defined splitting. This indicates the onset of internal fields with some field distribution, evidencing the onset of short range magnetic ordering as could be expected from the diluted Eu sublattice. The least square fit of spectra at 4.2 K shows a hyperfine field of 11 Tesla at Eu nucleus. This value of hyperfine field is substantially smaller (nearly half) than that for EuFe$_{2}$As$_{2}
 $ (B$_{hf}$ = 26 Tesla) which is due to doping of 50\% Europium with non magnetic potassium \cite{23}.

$^{57}$Fe M\"{o}ssbauer spectra also reveal important information about this system (Fig. 6). At room temperature we see a single unsplit line with a width of 0.32 mm/s confirming absence of un-reacted iron in the sample. The line width does not show any change at 180 K or at lower temperature down to 20 K suggesting absence of SDW transition which was present in the parent compound EuFe$_{2}$As$_{2}$ at 190 K. Only below 15 K we see a marginal increase of line width (0.5 mm/s at 4.2K) due to transferred field from Eu-ordering.

\section{Conclusion}

To summarize our combined investigation of the magnetic susceptibility, electrical resistivity and M\"{o}ssbauer spectroscopy  clearly establish coexistence of Eu short range magnetic order and superconductivity in Eu$_{0.5}$K$_{0.5}$Fe$_{2}$As$_{2}$ sample. Superconductivity in this sample occurs below 33 K and magnetic order due to Eu-moments occurs below 13 K. The features of magnetic susceptibility in the temperature range 5-15 K is similar to that found in HoNi$_{2}$B$_{2}$C which was attributed to the intense competition between Ho-magnetic order and superconductivity.

\section*{Acknowledgement}
Financial support from BRNS and IIT Kanpur is acknowledged.

\end{document}